\documentclass[preprint]{elsarticle}

\usepackage{hyperref,bm}
\usepackage{color}
\usepackage{ulem}

\usepackage{amsmath,amssymb}
\usepackage{booktabs}

\journal{Journal of \LaTeX\ Templates}

\begin{document}

\begin{frontmatter}

\title{
Skyrmion crystals in centrosymmetric triangular magnets under hexagonal and trigonal single-ion anisotropy
}

\author{Satoru Hayami}
\address{Department of Applied Physics, The University of Tokyo, Bunkyo, Tokyo 113-8656, Japan}
\ead{hayami@ap.t.u-tokyo.ac.jp}

\begin{abstract}
We theoretically report an instability toward a skyrmion crystal in centrosymmetric magnets under hexagonal and trigonal single-ion anisotropy. 
The results are obtained for a minimal spin model with a crystal-dependent single-ion anisotropy on a triangular lattice by performing simulated annealing. 
By constructing magnetic phase diagrams in the presence of six different types of single-ion anisotropy while changing the amplitudes of external magnetic field and single-ion anisotropy in a systematic way, we find that the hexagonal and trigonal single-ion anisotropy becomes a source of the skyrmion crystal depending on the magnetic-field direction. 
We show that the skyrmion crystal is stabilized by the uniaxial-type or trigonal-type single-ion anisotropy under the out-of-plane magnetic field, while it is stabilized by the hexagonal-type inplane single-ion anisotropy under the inplane magnetic field. 
We also find various multiple-$Q$ spin states depending on the types of the single-ion anisotropy. 
\end{abstract}


\end{frontmatter}


\section{Introduction}
\label{sec:Introduction}

A magnetic skyrmion, which is characterized by a topologically nontrivial swirling spin texture, has been the subject of considerable interest in both theory and experiment in condensed matter physics~\cite{Bogdanov89,Bogdanov94,rossler2006spontaneous,nagaosa2013topological,Tokura_doi:10.1021/acs.chemrev.0c00297,hayami2021topological}. 
A periodic array of the magnetic skyrmions, referred to as the skyrmion crystal (SkX), was originally found in cubic chiral magnets without spatial inversion symmetry, such as MnSi~\cite{ishikawa1976helical,Muhlbauer_2009skyrmion,Neubauer_PhysRevLett.102.186602}, Fe$_{1-x}$Co$_x$Si~\cite{yu2010real,beille1983long}, FeGe~\cite{yu2011near,lebech1989magnetic}, and Cu$_2$OSeO$_3$~\cite{seki2012observation,Adams2012,Seki_PhysRevB.85.220406}, and subsequently, it was also found in other noncentrosymmetric polar magnets with square and trigonal symmetry~\cite{kezsmarki_neel-type_2015,nayak2017discovery,Fujima_PhysRevB.95.180410,bordacs2017equilibrium,Kurumaji_PhysRevLett.119.237201,peng2020controlled} and surface~\cite{heinze2011spontaneous,romming2013writing}. 
More recently, the SkX was observed in centrosymmetric magnets, such as Gd$_2$PdSi$_3$~\cite{Saha_PhysRevB.60.12162,kurumaji2019skyrmion,sampathkumaran2019report,Hirschberger_PhysRevB.101.220401,Kumar_PhysRevB.101.144440,spachmann2021magnetoelastic}, Gd$_3$Ru$_4$Al$_{12}$~\cite{chandragiri2016magnetic,Nakamura_PhysRevB.98.054410,hirschberger2019skyrmion,Hirschberger_10.1088/1367-2630/abdef9}, and GdRu$_2$Si$_2$~\cite{khanh2020nanometric,Yasui2020imaging,khanh2022zoology}. 
These findings of the SkXs in a wide range of materials with distinct lattice and electronic structures indicate various stabilization mechanisms. 

In noncentrosymmetric skyrmion-hosting magnets, it was well recognized that the Dzyaloshinskii-Moriya (DM) interaction~\cite{dzyaloshinsky1958thermodynamic,moriya1960anisotropic}, which arises from the relativistic spin-orbit coupling without spatial inversion symmetry, plays an essential role to stabilize the SkX under an external magnetic field~\cite{rossler2006spontaneous,Yi_PhysRevB.80.054416,Binz_PhysRevLett.96.207202,Binz_PhysRevB.74.214408,Hayami_PhysRevLett.121.137202}, although the axial anisotropy affects the stability range of the SkX phase~\cite{PhysRevB.94.014406,PhysRevB.96.214413,PhysRevB.102.104407}. 
In addition, the interplay between the DM and higher-spin interactions gives rise to not only the SkX with the small magnetic period in the Fe/Ir interface~\cite{heinze2011spontaneous} and EuPtSi~\cite{kakihana2018giant,kaneko2019unique,tabata2019magnetic,kakihana2019unique,hayami2021field} but also other topological spin crystals, such as a hedgehog lattice in MnSi$_{1-x}$Ge$_{x}$~\cite{tanigaki2015real,kanazawa2017noncentrosymmetric,fujishiro2019topological,Kanazawa_PhysRevLett.125.137202,grytsiuk2020topological,Okumura_PhysRevB.101.144416,Mendive-Tapia_PhysRevB.103.024410,Kato_PhysRevB.104.224405} and a meron-antimeron crystal~\cite{yu2018transformation,Hayami_PhysRevB.104.094425}. 
In this way, rich topological magnetic textures in noncentrosymmetric magnets are brought about by the synergy among several types of interactions: the ferromagnetic exchange interaction, the DM interaction, the axial anisotropy, and the multiple-spin interaction, where the former two interactions lead to the spiral modulation of spins, while the last multi-spin interaction tends to superpose a spiral state resulting in an exotic multiple-$Q$ state with the small magnetic period. 
Furthermore, recent theoretical studies have shown that the SkX can also appear in the centrosymmetric systems with the local-type DM interaction~\cite{hayami2021skyrmion,lin2021skyrmion}. 

Meanwhile, the DM interaction is not necessarily for the emergence of the multiple-$Q$ states including the SkX~\cite{batista2016frustration,hayami2021topological}. 
The pioneering work by Okubo, Chung, and Kawamura has clarified that the SkX appears at finite temperatures in the triangular-lattice Heisenberg model with the competing ferromagnetic and antiferromagnetic exchange interactions but without the DM interaction~\cite{Okubo_PhysRevLett.108.017206}. 
Then, the ground-state SkX has been clarified by additionally taking into account the uniaxial single-ion anisotropy~\cite{leonov2015multiply,Lin_PhysRevB.93.064430,Hayami_PhysRevB.93.184413,Lin_PhysRevLett.120.077202,Hayami_PhysRevB.103.224418}, two-ion magnetic anisotropy~\cite{amoroso2020spontaneous,Wang_PhysRevB.103.104408,amoroso2021tuning}, dipolar interaction~\cite{Utesov_PhysRevB.103.064414,utesov2021mean} and nonmagnetic impurity~\cite{Hayami_PhysRevB.94.174420}. 
Furthermore, another mechanism to stabilize the SkX has been established based on the spin-charge coupling in itinerant magnets~\cite{Ozawa_PhysRevLett.118.147205,Hayami_PhysRevB.95.224424,hayami2021topological,hayami2021phase,wang2021skyrmion}. 
Although the long-range nature of the interaction like the Ruderman-Kittel-Kasuya-Yosida (RKKY) interaction~\cite{Ruderman,Kasuya,Yosida1957} in itinerant magnets is different from the short-range one in frustrated insulating magnets, an effective multi-spin interaction arising from the Fermi surface instability gives rise to the ground-state SkX even without the DM interaction~\cite{Hayami_PhysRevB.95.224424}. 
Similar to the frustrated magnets, it was shown that the effect of single-ion anisotropy~\cite{Hayami_PhysRevB.99.094420,hayami2020multiple,Wang_PhysRevLett.124.207201,Hayami_10.1088/1367-2630/ac3683}, two-ion anisotropy~\cite{Hayami_doi:10.7566/JPSJ.89.103702,yambe2021skyrmion,Hayami_PhysRevB.103.024439,Hayami_PhysRevB.103.054422,hayami2022multiple,yambe2022effective}, circularly polarized microwave field~\cite{Eto_PhysRevB.104.104425}, and thermal fluctuations~\cite{Mitsumoto_PhysRevB.104.184432,mitsumoto2021skyrmion} enhance the stability of the SkX. 
It was also shown that the introduction of the DM or antisymmetric spin-orbit interaction in itinerant magnets leads to the SkX formation~\cite{Hayami_PhysRevLett.121.137202,Nikoli_PhysRevB.103.155151,Kathyat_PhysRevB.102.075106,mukherjee2021antiferromagnetic,Kathyat_PhysRevB.103.035111}.

The above theoretical investigations have revealed that the SkXs are stabilized under various interactions and magnetic anisotropy in both insulating and metallic systems. 
In the present study, to further explore the SkX-hosting physical systems based on the microscopic model, we consider the effect of the crystal-dependent single-ion anisotropy in the hexagonal and trigonal systems systematically. 
By analyzing a simplified spin model including the exchange interaction, single-ion anisotropy, and magnetic field, and performing simulated annealing, we show that any type of single-ion anisotropy can be a source of the SkX. 
In addition to the uniaxial anisotropy, the trigonal anisotropy tends to stabilize the SkX in an out-of-plane magnetic field.
Moreover, we find that the interplay between the hexagonal inplane single-ion anisotropy and the inplane magnetic field also induces the SkX. 
Besides, we find a variety of multiple-$Q$ states distinct from the SkX, which have not been reported so far. 
Our results indicate that the single-ion anisotropy is another key ingredient to induce the multiple-$Q$ states including the SkX in centrosymmetric magnets. 

The rest of the paper is organized as follows.
In Sec.~\ref{sec:Model and method}, we introduce a spin model with the hexagonal and trigonal single-ion anisotropy
and outline the simulated annealing.  
We discuss the instability toward the SkX in Sec.~\ref{sec:Results}. 
We systematically show the magnetic phase diagrams under the uniaxial-type single-ion anisotropy, trigonal-type single-ion anisotropy, and hexagonal inplane single-ion anisotropy.  
A summary of results is given in Sec.~\ref{sec:Summary}. 

\section{Model and method}
\label{sec:Model and method}

\begin{table*}
\caption{
The relation between the single-ion anisotropy (SIA) $-A_{lm} \mathcal{O}^{(lm)}_{i}$ in Eq.~(\ref{eq:Ham_loc}) and the spherical harmonics $Y_{lm}$ and $Y^*_{lm}$ under the hexagonal and trigonal point groups;  $f^{c3\phi}_i=S_i^x[ (S_i^x)^2-3(S_i^y)^2]$ and $f^{s3\phi}_i=S_i^y   [3 (S_i^x)^2-(S_i^y)^2]$. 
Note that $D_{\rm 6h}$ and $D_{\rm 3d}$ have nonzero $A_{66}$ instead of $A_{66'}$, while $C_{\rm 3 i}$ and $C_{\rm 6h}$ have both $A_{lm}$ and $A_{lm'}$. 
The appearance of the SkX in an external field in each single-ion anisotropy is also shown, where the results in (43), (63), and (66) are the same as those in (43)', (63)', and (66)' by replacing the spin component appropriately. 
It is noted that the SkX can be stabilized in an inplane field in the case of (66)' [or (66)]. 
}
\label{tab}
\centering
\begin{tabular}{cccccccccccccc|ccc|ccc} \hline\hline
SIA ($-A_{lm} \mathcal{O}^{(lm)}_{i}$) & $Y_{lm}, Y^*_{lm}$ & point group & SkX in a $z$ field  \\  \hline
$-A_{20}(S_i^z)^2$  &  $(20)$ & $D_{\rm 6h}, C_{\rm 6h}, D_{\rm 3d}, C_{\rm 3 i}$ & $A_{20}>0$\\
$-A_{40}(S_i^z)^4$  &  $(40)$ & $D_{\rm 6h}, C_{\rm 6h}, D_{\rm 3d}, C_{\rm 3 i}$ & $A_{40}>0$\\
$-A_{60}(S_i^z)^6$  &  $(60)$ & $D_{\rm 6h}, C_{\rm 6h}, D_{\rm 3d}, C_{\rm 3 i}$ & $A_{60}>0$\\
$-A_{43'} S_i^z f^{s3\phi}_i$  &  $(43)'$ & $D_{\rm 3d}, C_{\rm 3 i}$ & $|A_{43'}|>0$\\
$-A_{63'} (S_i^z)^3 f^{s3\phi}_i$  &  $(63)'$ & $D_{\rm 3d}, C_{\rm 3 i}$ & $|A_{63'}|>0$\\
$-A_{66'}  f^{c3\phi}_i f^{s3\phi}_i$ & $(66)'$  &  $D_{\rm 6h}, C_{\rm 6h}, D_{\rm 3d}, C_{\rm 3 i}$ & No \\
\hline\hline
\end{tabular}
\end{table*}

Let us start from a Heisenberg model with a classical spin on a two-dimensional triangular lattice. 
Hereafter, we take the lattice constant of the triangular lattice as unity. 
The Hamiltonian is given by 
\begin{eqnarray}
\label{eq:Ham}
\mathcal{H}&=& \mathcal{H}_{\rm ex}+\mathcal{H}_{\rm loc}, \\
\label{eq:Ham_ex}
\mathcal{H}_{\rm ex} &=& \sum_{\langle i  j\rangle }J_{ij} \bm{S}_i \cdot \bm{S}_j , \\
\label{eq:Ham_loc}
\mathcal{H}_{\rm loc} &=& -\sum_{i} \sum_{lm} A_{lm} \mathcal{O}^{(lm)}_{i} - H \sum_i S_i^z,  
\end{eqnarray}
where $\bm{S}_i$ is the classical localized spin with $|\bm{S}_i|=1$. 
The total Hamiltonian $\mathcal{H}$ consists of the exchange Hamiltonian $\mathcal{H}_{\rm ex}$ and the local Hamiltonian $\mathcal{H}_{\rm loc}$. 
The exchange Hamiltonian $\mathcal{H}_{\rm ex}$ includes the interactions between further neighboring spins, e.g., $J_{ij}=J_1$ ($J_2$) represents the (next-)nearest-neighbor interaction. 
The local Hamiltonian $\mathcal{H}_{\rm loc}$ in Eq.~(\ref{eq:Ham_loc}) includes the single-ion magnetic anisotropy in the first term and the Zeeman coupling in an external magnetic field in the second term. 
We here incorporate the effect of the single-ion anisotropy arising from the hexagonal and trigonal crystal symmetry; $\mathcal{O}^{(lm)}_{i}$ is represented by the spin product, which is anisotropic in spin space and $A_{lm}$ is the coefficient of $\mathcal{O}^{(lm)}_{i}$. 
The anisotropy of $\mathcal{O}^{(lm)}_{i}$ in spin space is characterized by the spherical Harmonics $Y_{lm}$ with the integers $l$ and $m$ with $-l \leq m\leq l$.  
When considering the centrosymmetric hexagonal and trigonal point groups, i.e., $D_{\rm 6h}$, $C_{\rm 6h}$, $D_{\rm 3d}$, and $C_{\rm 3 i}$, nonzero $\mathcal{O}^{(lm)}_{i}$ up to $l=6$ is given by 
\begin{eqnarray}
\label{eq:O20}
\mathcal{O}^{(20)}_i&=&(S_i^z)^2, \\
\label{eq:O40}
\mathcal{O}^{(40)}_i&=&(S_i^z)^4, \\
\label{eq:O60}
\mathcal{O}^{(60)}_i&=&(S_i^z)^6, \\
\label{eq:O43}
\mathcal{O}^{(43)}_i&=&S_i^z S_i^x[ (S_i^x)^2-3(S_i^y)^2], \\
\label{eq:O43'}
\mathcal{O}^{(43')}_i&=&S_i^z S_i^y[ 3(S_i^x)^2-(S_i^y)^2], \\
\label{eq:O63}
\mathcal{O}^{(63)}_i&=&(S_i^z)^3 S_i^x[ (S_i^x)^2-3(S_i^y)^2], \\
\label{eq:O63'}
\mathcal{O}^{(63')}_i&=&(S_i^z)^3 S_i^y[ 3(S_i^x)^2-(S_i^y)^2], \\
\label{eq:O66}
\mathcal{O}^{(66)}_i&=&[(S_i^x)^6-15 (S_i^x)^2 (S_i^y)^2\{(S_i^x)^2-(S_i^y)^2\}-(S_i^y)^6], \\
\label{eq:O66'}
\mathcal{O}^{(66')}_i&=&S_i^x S_i^y[ (S_i^x)^2-3(S_i^y)^2][ 3(S_i^x)^2-(S_i^y)^2], 
\end{eqnarray}
where $(lm)$ and $(lm')$ correspond to $(-1)^m(Y_{lm}+Y^*_{lm})/\sqrt{2}$ and $(-1)^m(Y_{lm}-Y^*_{lm})/\sqrt{2}i$, respectively~\cite{hutchings1964point,Kusunose_JPSJ.77.064710,Hayami_PhysRevB.98.165110}. 
From the symmetry viewpoint, the nonzero single-ion anisotropy for $D_{\rm 6h}$, $C_{\rm 6h}$, $D_{\rm 3d}$, and $C_{\rm 3 i}$ is given by  ($\mathcal{O}^{(20)}_i$, $\mathcal{O}^{(40)}_i$, $\mathcal{O}^{(60)}_i$, $\mathcal{O}^{(66)}_i$), 
($\mathcal{O}^{(20)}_i$, $\mathcal{O}^{(40)}_i$, $\mathcal{O}^{(60)}_i$, $\mathcal{O}^{(66)}_i$, $\mathcal{O}^{(66')}_i$), 
($\mathcal{O}^{(20)}_i$, $\mathcal{O}^{(40)}_i$, $\mathcal{O}^{(43')}_i$, $\mathcal{O}^{(60)}_i$, $\mathcal{O}^{(63')}_i$, $\mathcal{O}^{(66)}_i$), and 
($\mathcal{O}^{(20)}_i$, $\mathcal{O}^{(40)}_i$, $\mathcal{O}^{(43)}_i$, $\mathcal{O}^{(43')}_i$, $\mathcal{O}^{(60)}_i$, $\mathcal{O}^{(63)}_i$, $\mathcal{O}^{(63')}_i$, $\mathcal{O}^{(66)}_i$, $\mathcal{O}^{(66')}_i$), respectively. 
The appearance of $\mathcal{O}^{(l3)}_i$ and $\mathcal{O}^{(l3')}_i$ in the trigonal point group is owing to the lacking of the sixfold rotational symmetry. 
We summarize the correspondence among the single-ion anisotropy, spherical harmonics, and point groups in Table~\ref{tab}. 

In the following, we investigate the effect of $\mathcal{O}^{(lm)}_{i}$ on the stabilization of the SkX in the ground state on the triangular lattice. 
In order to find the low-energy state and the instability toward the SkX under the single-ion anisotropy in Eqs.~(\ref{eq:O20})-(\ref{eq:O66'}) in a systematic and efficient manner, we introduce a simplified spin model derived from $\mathcal{H}$ as follows: 
\begin{equation}
\label{eq:Hameff}
\tilde{\mathcal{H}}= -\sum_{\nu} J_{\bm{Q}_\nu} \bm{S}_{\bm{Q}_{\nu}} \cdot \bm{S}_{-\bm{Q}_{\nu}}+\mathcal{H}_{\rm loc}, 
\end{equation}
where $\mathcal{H}_{\rm ex}$ is replaced with the first term in Eq.~(\ref{eq:Hameff}); $\bm{S}_{\bm{Q}_\nu}$ is the Fourier transform of $\bm{S}_i$ with the wave vector ${\bm{Q}_\nu}$. 
The first term in Eq.~(\ref{eq:Hameff}) stands for the simplified exchange interaction by extracting the dominant $\bm{q}$ contribution in the Fourier transform of $\mathcal{H}_{\rm ex}$. 
In other words, we only consider the interactions in the $\bm{Q}_\nu$ channel and ignore the contributions from the other wave vectors.  
This treatment is justified when considering the low-temperature (ground) state, since the magnetic ordered states are characterized by the single-$Q$ or multiple-$Q$ spin density waves consisting of the wave vector $\bm{Q}_{\nu}$ that gives the lowest energy of the spin Hamiltonian in Eq.~(\ref{eq:Ham})~\cite{leonov2015multiply,Hayami_PhysRevB.103.224418}. 
As for $\bm{Q}_{\nu}$, we consider the contributions from six $\bm{Q}_{\nu}$ to satisfy the threefold rotational symmetry of the triangular lattice: $\bm{Q}_1=(Q,0)$, $\bm{Q}_2=(-Q/2,\sqrt{3}Q/2)$, and $\bm{Q}_3=(-Q/2,-\sqrt{3}Q/2)$ with $J_{\bm{Q}_\nu}=\tilde{J}$ and $Q=2\pi/9$. 
Such a situation can be realized by considering the ferromagnetic nearest-neighbor exchange interaction $J_1<0$ and antiferromagnetic third-nearest-neighbor interaction $J_3>0$ while keeping $J_3/|J_1|>1/4$ ($J_3/|J_1|\simeq 0.3025$ for $Q=2\pi/9$). 
We here neglect the contributions from the higher harmonics like $\bm{Q}_1+\bm{Q}_2$ and $2\bm{Q}_1$ and $\bm{q}=\bm{0}$ components in the interactions for simplicity, although the former contribution sometimes play a role in the SkX formation~\cite{hayami2022multiple}. 
We take $\tilde{J}=1$ as the energy unit of the model in Eq.~(\ref{eq:Hameff}). 

We perform the simulated annealing for the spin model in Eq.~(\ref{eq:Hameff}) to examine the low-temperature spin states including the SkX following the manner in Ref.~\cite{Hayami_PhysRevB.103.224418}.  
The simulated annealing is carried out from high temperatures $T=0.1$-$1.0$ down to low temperatures $T=0.01$ with a rate $\alpha=0.99999$-$0.999999$; the $n$th-step temperature $T_n$ is given by $T_{n+1}=\alpha T_n$. 
In each temperature, the simulations are performed for the system size with total spins $N=36^2$ under the periodic boundary condition by using the standard Metropolis local updates. 
The $10^5$-$10^6$ Monte Carlo sweeps are taken for measurements. 
Although the simulations are performed by starting from a random spin configuration from high temperatures, we also start the simulations from the spin configurations obtained at low temperatures near the phase boundary where several magnetic states have almost the same energy.

To identify magnetic phases, we calculate the inplane and out-of-plane spin structure factors as 
\begin{eqnarray}
S_s^{xy}(\bm{q})&=& \frac{1}{N} \sum_{i,j} (S^{x}_i S^{x}_j+S^{y}_i S^{y}_j) e^{i\bm{q}\cdot (\bm{r}_i-\bm{r}_j)}, \\
S_s^{z}(\bm{q})&=& \frac{1}{N} \sum_{i,j} S^{z}_i S^{z}_j e^{i\bm{q}\cdot (\bm{r}_i-\bm{r}_j)}, 
\end{eqnarray}
and the net magnetization along the field direction as $M^z=(1/N)\sum_i S_i^z$. 
Then, the magnetic moments with wave vector $\bm{Q}_\nu$ are given by $m^{xy}_{\bm{Q}_{\eta}}=\sqrt{S_s^{xy}(\bm{Q}_{\eta})/N}$ and $m^{z}_{\bm{Q}_{\eta}}=\sqrt{S_s^{z}(\bm{Q}_{\eta})/N}$. 
We also compute the scalar chirality 
\begin{equation}
\chi_0 = \frac{1}{N} \sum_{\mu} \sum_{\bm{R} \in \mu} \bm{S}_{i} \cdot (\bm{S}_j \times \bm{S}_k),
\end{equation}
where $\bm{R}$ stand for the position vectors at the centers of triangles consisting of three sites $i$, $j$, and $k$ in the counterclockwise order, and $\mu=(u,d)$ stand for upward and downward triangles, respectively. 
The nonzero $\chi_0$ implies the emergence of the SkX with the quantized skyrmion number of $\pm 1$.

\section{Results}
\label{sec:Results}

We examine the effect of single-ion magnetic anisotropy in Eqs.~(\ref{eq:O20})-(\ref{eq:O66'}) that appears in centrosymmetric hexagonal and trigonal point groups on the SkX by performing the simulated annealing. 
In the following, we discuss the effect of single-ion anisotropy one by one by categorizing into the three types: uniaxial-type anisotropy in Sec.~\ref{sec:Uniaxial anisotropy}, trigonal-type anisotropy in Sec.~\ref{sec:Trigonal anisotropy}, and hexagonal-type inplane anisotropy in Sec.~\ref{sec:Hexagonal inplane anisotropy}.

\subsection{Uniaxial anisotropy}
\label{sec:Uniaxial anisotropy}

\begin{figure}[t!]
\begin{center}
\includegraphics[width=0.6 \hsize ]{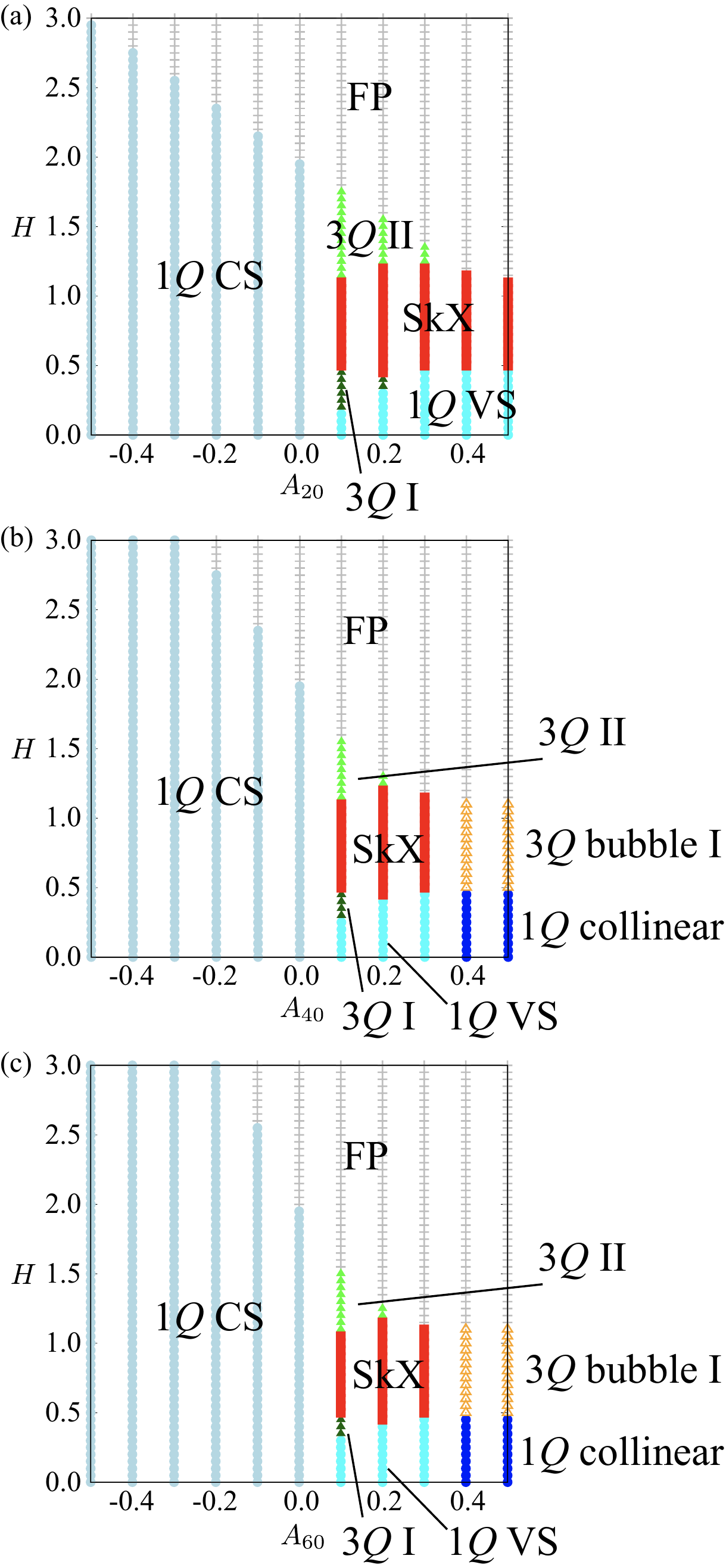} 
\caption{
\label{fig:PD_Bz}
Magnetic phase diagram under the uniaxial single-ion anisotropy; (a) $A_{20}$, (b) $A_{40}$, and (c) $A_{60}$. 
The positive $A_{20}$, $A_{40}$, and $A_{60}$ stand for the easy-axis anisotropy, while the negative ones stand for the easy-plane anisotropy. 
The magnetic field $H$ is applied along the $z$ direction.
``SkX", ``VS", ``CS", and ``FP" represent the skyrmion crystal, vertical spiral, conical spiral, and the fully-polarized states, respectively. 
}
\end{center}
\end{figure}

\begin{figure}[t!]
\begin{center}
\includegraphics[width=0.8 \hsize ]{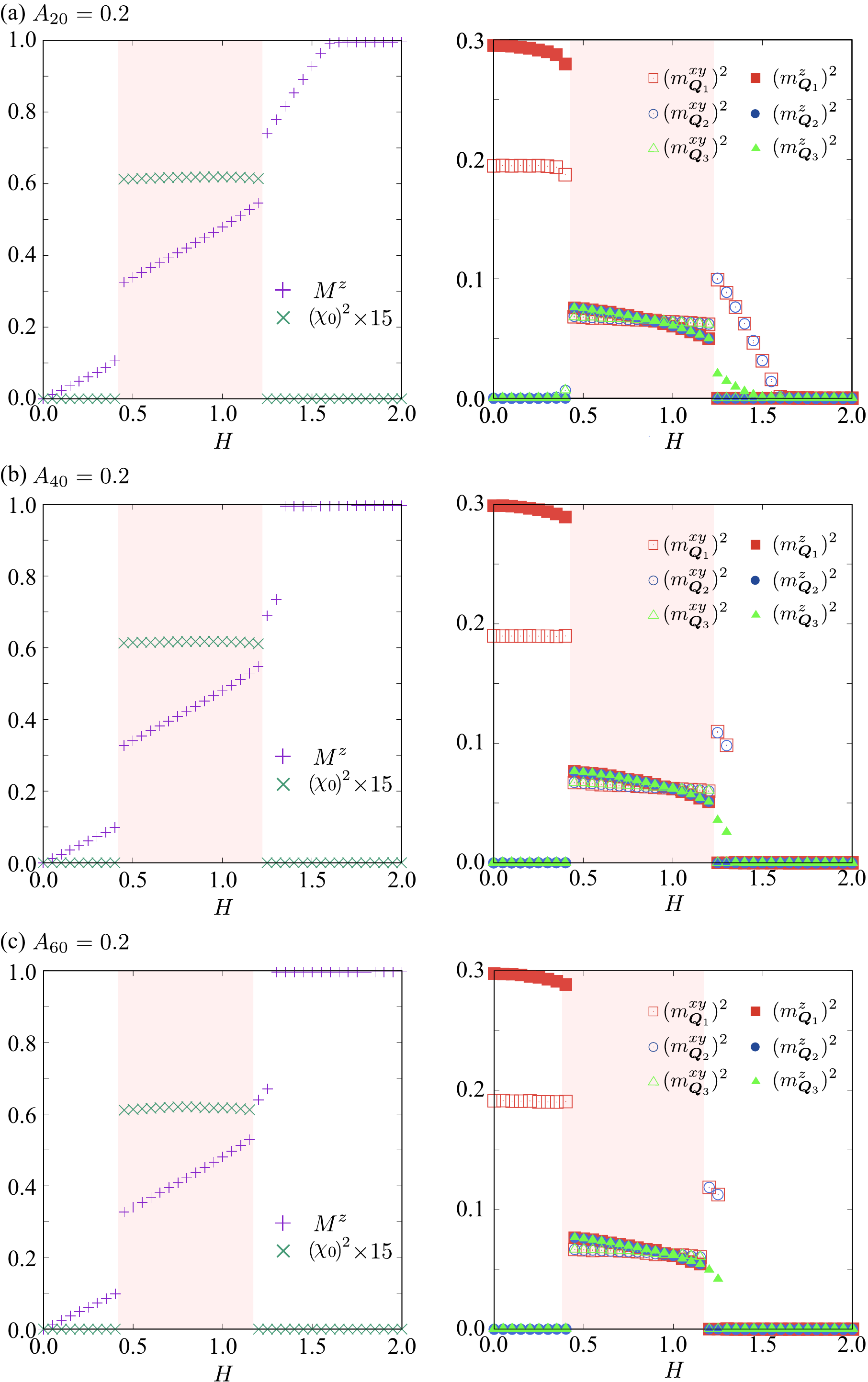} 
\caption{
\label{fig:Bz_Mq}
$H$ dependence of (left panel) $M^z$ and $(\chi_0)^2$ and (right panel) $(m_{\bm{Q}_\eta}^{xy})^2$ and $(m_{\bm{Q}_\eta}^{z})^2$ for (a) $A_{20}=0.2$, (b) $A_{40}=0.2$, and (c) $A_{60}=0.2$. 
The magnetic field $H$ is applied along the $z$ direction.
The region drawn by red represents the SkX phase. 
}
\end{center}
\end{figure}

\begin{figure}[t!]
\begin{center}
\includegraphics[width=0.8 \hsize ]{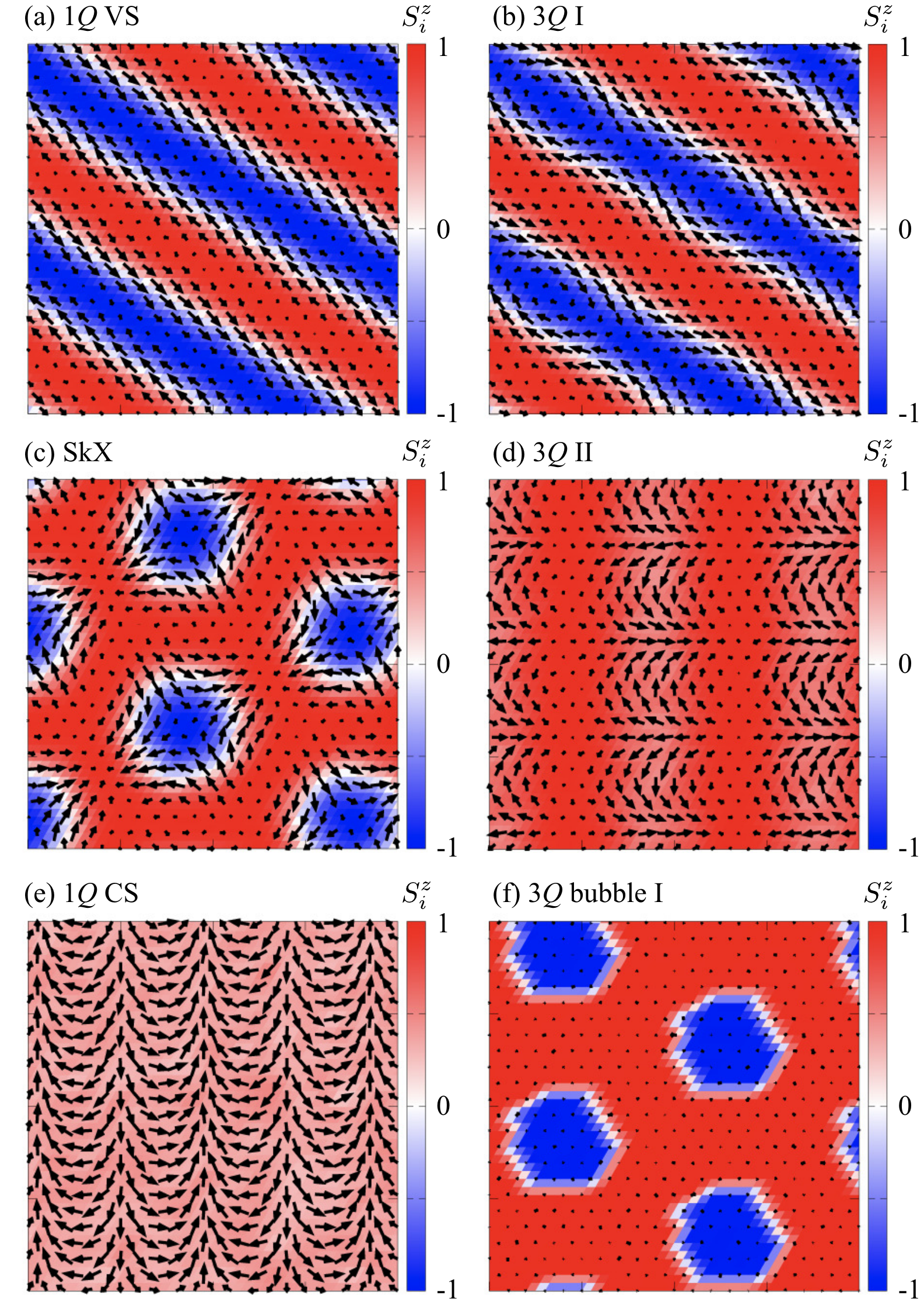} 
\caption{
\label{fig:Bz_spin}
Real-space spin configurations of (a) the single-$Q$ vertical spiral (1$Q$ VS) state at $A_{20}=0.2$ and $H=0.2$, (b) the triple-$Q$ (3$Q$) I state at  $A_{20}=0.2$ and $H=0.4$, (c) the SkX at  $A_{20}=0.2$ and $H=0.8$, (d) the 3$Q$ II state at  $A_{20}=0.2$ and $H=1.3$, (e) the 1$Q$ conical spiral (CS) state at  $A_{20}=-0.2$ and $H=0.9$, and (f) the 3$Q$ bubble I state at  $A_{40}=0.4$ and $H=0.8$. 
The contour shows the $z$ component of the spin moment, and the arrows represent the $xy$ components. 
}
\end{center}
\end{figure}

We discuss the results in the presence of uniaxial-type single-ion anisotropy, namely, $A_{20}$, $A_{40}$, and $A_{60}$, which has been studied in various magnetic systems~\cite{Liu_PhysRevB.60.12893,PhysRevB.79.014415,si2022ferrimagnetism}. 
Although some of the results for $A_{20}$ have already been reported in the literatures~\cite{leonov2015multiply,Lin_PhysRevB.93.064430,Hayami_PhysRevB.93.184413,Hayami_PhysRevB.103.054422}, we discuss the effect of $A_{20}$ again to make the present paper self-contained and to compare the results with $A_{40}$ and $A_{60}$. 
Figure~\ref{fig:PD_Bz}(a) shows the low-temperature phase diagram while changing $A_{20}$ and $H$. 
The positive (negative) $A_{20}$ represents the easy-axis (easy-plane) single-ion anisotropy. 
The phase diagram becomes asymmetric in terms of $A_{20}$; the multiple-$Q$ and SkX phases in addition to the single-$Q$ vertical spiral phase appear for $A_{20}>0$, while only the single-$Q$ conical spiral state appears for $A_{20}<0$. 
The obtained phase diagram is consistent with those of previous studies in Refs.~\cite{leonov2015multiply,Lin_PhysRevB.93.064430,Hayami_PhysRevB.93.184413,Hayami_PhysRevB.103.054422}. 

We show the $H$ dependence of spin- and chirality-related quantities in the case of $A_{20}=0.2$ in Fig.~\ref{fig:Bz_Mq}(a), where there are four magnetic phases in addition to the fully-polarized state. 
Here and hereafter, we show the results in each ordered state by appropriately sorting $m^{xy}_{\bm{Q}_\eta}$ and $m^{z}_{\bm{Q}_\eta}$ for better readability. 
At zero field, the single-$Q$ vertical spiral state appears, where the relation $m^{z}_{\bm{Q}_1}> m^{xy}_{\bm{Q}_1}$ is owing to the presence of the easy-axis single-ion anisotropy. 
While increasing $H$, the uniform magnetization $M^z$ linearly increases and $m^{z}_{\bm{Q}_1}$ gradually decreases, as shown in Fig.~\ref{fig:Bz_Mq}(a). 
The real-space spin configuration obtained by the simulated annealing is shown in Fig.~\ref{fig:Bz_spin}(a).  
Around $H \simeq 0.4$, the $\bm{Q}_2$ and $\bm{Q}_3$ components of magnetic moments, $m^{xy}_{\bm{Q}_2}$ and $m^{xy}_{\bm{Q}_3}$, become nonzero, which indicate the appearance of a triple-$Q$ state. 
We call this state a triple-$Q$ I state, whose spin configuration is shown in Fig.~\ref{fig:Bz_spin}(b). 
Compared to the spin configurations in Figs.~\ref{fig:Bz_spin}(a) and \ref{fig:Bz_spin}(b), one finds that the vortex-like spin configuration appears in the region for $S_i^z <0$ in the triple-$Q$ I state, which reflects the triple-$Q$ nature of the inplane spin component. 
Accordingly, the triple-$Q$ I state exhibits the scalar chirality density waves along with the $\bm{Q}_2$ and $\bm{Q}_3$ directions, which results in the checkerboard-type modulations of the scalar chirality. 

While further increasing $H$, the triple-$Q$ I state turns into the SkX, where there is a jump of $M^z$, as shown in Fig.~\ref{fig:Bz_Mq}(a). 
The SkX is stabilized for $0.45 \lesssim H \lesssim 1.2$. 
The SkX is characterized by a superposition of three spiral states along the $\bm{Q}_1$, $\bm{Q}_2$, and $\bm{Q}_3$ directions with equal intensity, i.e., $m^{xy}_{\bm{Q}_1}=m^{xy}_{\bm{Q}_2}=m^{xy}_{\bm{Q}_3}$ and $m^{z}_{\bm{Q}_1}=m^{z}_{\bm{Q}_2}=m^{z}_{\bm{Q}_3}$. 
The real-space spin configuration shown in Fig.~\ref{fig:Bz_spin}(c) consists of a periodic array of the skyrmion cores, which are identified as the position with $S_i^z =-1$, in a triangular-lattice way. 
The skyrmion core tends to be located at the center of the triangle for small $A$ and large $H$, while that is located at the site for large $A$ and small $H$~\cite{Hayami_PhysRevResearch.3.043158}.
The SkX has a net scalar chirality $\chi_0$, as shown in the left panel of Fig.~\ref{fig:Bz_Mq}(a), which results in a quantized skyrmion number. 
In the case of Fig.~\ref{fig:Bz_spin}(c), the SkX has the skyrmion number of $1$, which corresponds to the anti-type SkX.
Owing to the spin rotational symmetry in the model in Eq.~(\ref{eq:Hameff}), the SkXs with the skyrmion number of $\pm 1$, i.e., the SkX with the negative skyrmion number and the anti-type SkX with the positive one, are degenerate. 
The degeneracy between the SkX and the anti-SkX is lifted by considering other anisotropic interactions, e.g., the bond-dependent anisotropic exchange interaction~\cite{Hayami_PhysRevB.103.054422}. 

The increase of $H$ in the SkX phase leads to the transition to the other triple-$Q$ state denoted as triple-$Q$ II state at $H \simeq 1.25$ with jumps of $M^z$ and $\chi_0$. 
The triple-$Q$ II state is characterized by $m^{xy}_{\bm{Q}_1}=m^{xy}_{\bm{Q}_2}$ and $m^{z}_{\bm{Q}_3}$ without the net scalar chirality. 
The spin configuration is shown in Fig.~\ref{fig:Bz_spin}(d). 
Similar to the triple-$Q$ I state, this state shows the chirality density waves along with the $\bm{Q}_3$ direction like the chiral stripe state found in the itinerant electron model~\cite{Ozawa_doi:10.7566/JPSJ.85.103703,yambe2020double}.
The triple-$Q$ II state turns into the fully-polarized state with the moments along the $z$ direction at $H\simeq 1.6$. 
Meanwhile, in the case of $A_{20}<0$, only the single-$Q$ conical spiral state with $m^{xy}_{\bm{Q}_1}$ but without $m^{z}_{\bm{Q}_1}$, whose spin configuration is shown in Fig.~\ref{fig:Bz_spin}(e), is stabilized except for the fully-polarized state, as shown in Fig.~\ref{fig:PD_Bz}(a). 

Figure~\ref{fig:PD_Bz}(b) shows the phase diagram under $A_{40}$. 
In addition to the six magnetic phases in Fig.~\ref{fig:PD_Bz}(a), two phases additionally appear in the large $A_{40}$ region: a single-$Q$ collinear state at low fields and a triple-$Q$ bubble I state at intermediate fields. 
The single-$Q$ collinear state corresponds to the single-$Q$ vertical spiral state without $m^{xy}_{\bm{Q}_1}$; the moments are along the $z$ direction.  
Similar to the correspondence between the single-$Q$ vertical spiral and collinear states, the triple-$Q$ bubble I state corresponds to the SkX without the $xy$ spin component~\cite{Hayami_PhysRevB.93.184413}. 
Indeed, the real-space spin configuration of the triple-$Q$ bubble I state is similar to that of the SkX, as shown in Figs.~\ref{fig:Bz_spin}(c) and \ref{fig:Bz_spin}(f). 
In contrast to the SkX, the amplitude of $m^{z}_{\bm{Q}_{\eta}}$ is slightly different from each other depending on the magnetic field, which might be owing to a fine balance among the energy gain by the exchange energy, single-ion anisotropy, and magnetic field. 
A similar triple-$Q$ bubble I state has been discussed in the other systems with the dipolar interaction~\cite{utesov2021mean} and spin-charge coupling~\cite{Hayami_10.1088/1367-2630/ac3683}. 
The tendency where the multiple-$Q$ bubble crystal is stabilized under the uniaxial anisotropy is common to the tetragonal-lattice case~\cite{Su_PhysRevResearch.2.013160,seo2021spin}. 

We show the phase diagram under $A_{60}$ in Fig.~\ref{fig:PD_Bz}(c).  
The results clearly represent that the overall phase diagram is common to the case under $A_{40}$ in Fig.~\ref{fig:PD_Bz}(b), although there is a small quantitative difference of spin- and chirality-related quantities between them shown in Figs.~\ref{fig:Bz_Mq}(b) and \ref{fig:Bz_Mq}(c). 
Thus, the uniaxial anisotropy gives a similar tendency of the SkX instability when considering its small contribution.

\subsection{Trigonal anisotropy}
\label{sec:Trigonal anisotropy}

\begin{figure}[t!]
\begin{center}
\includegraphics[width=0.6 \hsize ]{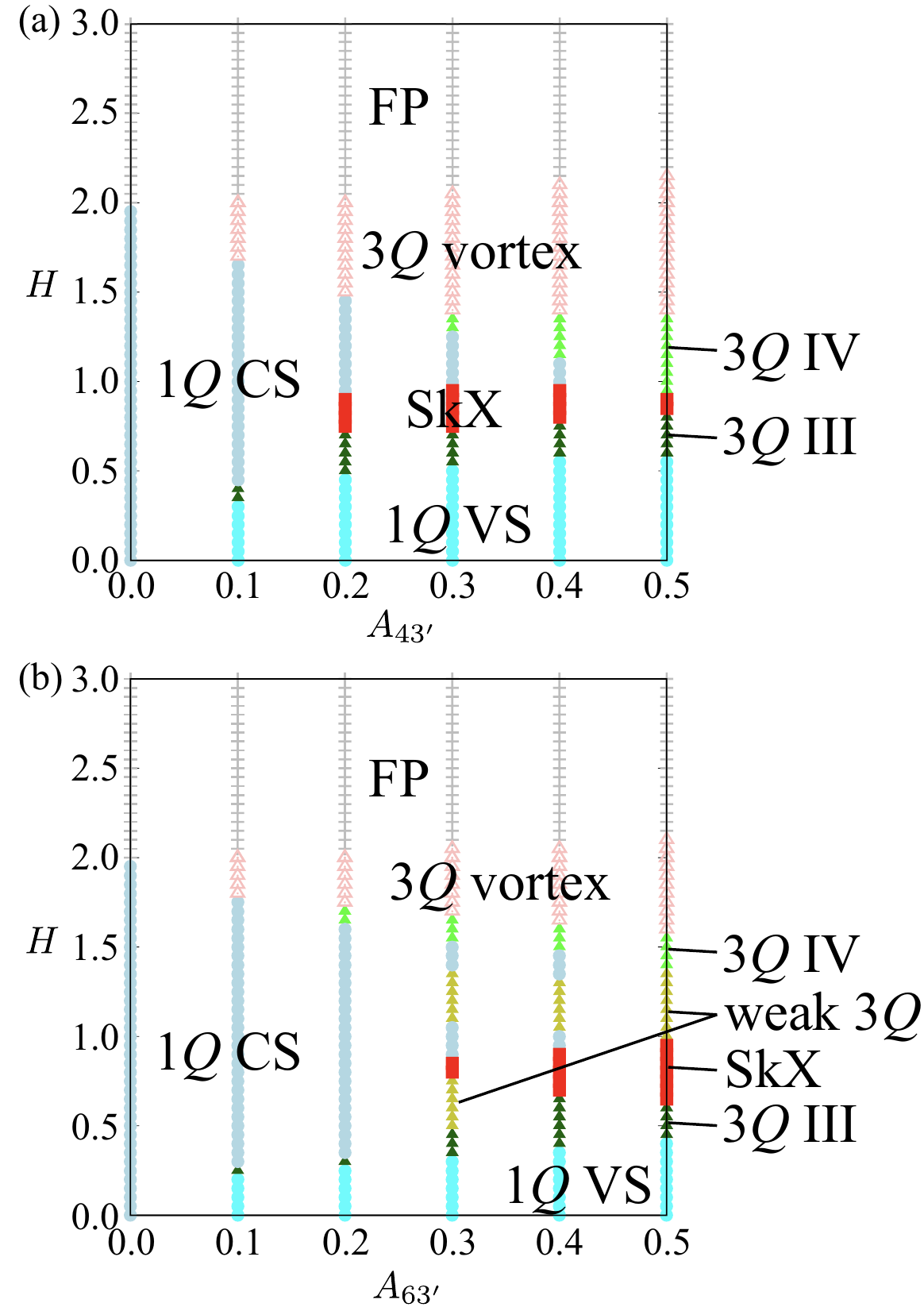} 
\caption{
\label{fig:PD_Byzx}
Magnetic phase diagram under the trigonal single-ion anisotropy; (a) $A_{43'}$ and (b) $A_{63'}$. 
The magnetic field $H$ is applied along the $z$ direction. 
``SkX", ``VS", ``CS", and ``FP" represent the skyrmion crystal, vertical spiral, conical spiral, and the fully-polarized states, respectively. 
}
\end{center}
\end{figure}

\begin{figure}[t!]
\begin{center}
\includegraphics[width=0.8 \hsize ]{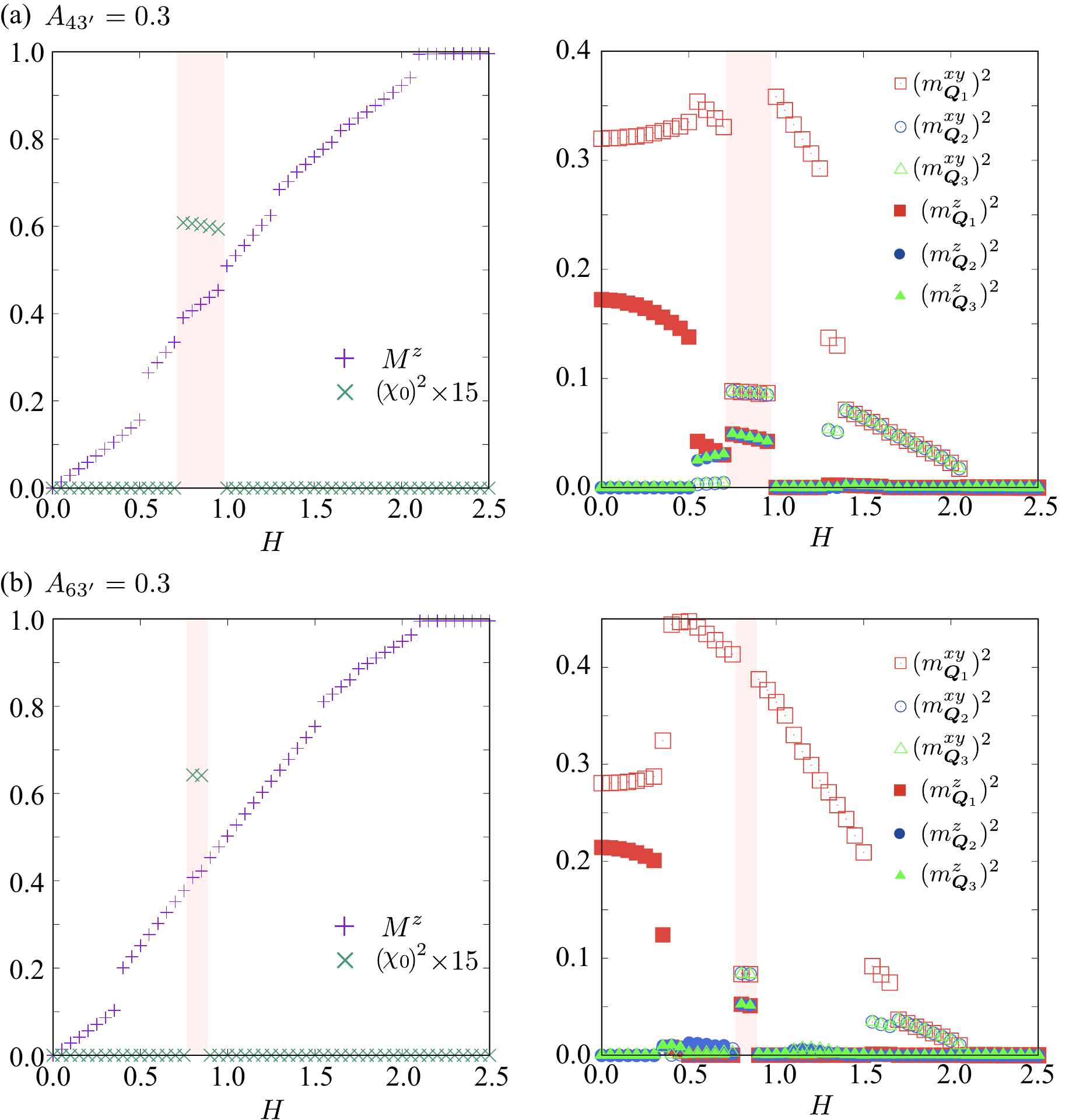} 
\caption{
\label{fig:Byzx_Mq}
$H$ dependence of (left panel) $M^z$ and $(\chi_0)^2$ and (right panel) $(m_{\bm{Q}_\eta}^{xy})^2$ and $(m_{\bm{Q}_\eta}^{z})^2$ for (a) $A_{43'}=0.3$ and (b) $A_{63'}=0.3$. 
The magnetic field $H$ is applied along the $z$ direction.
The region drawn by red represents the SkX phase. 
}
\end{center}
\end{figure}

\begin{figure}[t!]
\begin{center}
\includegraphics[width=0.8 \hsize ]{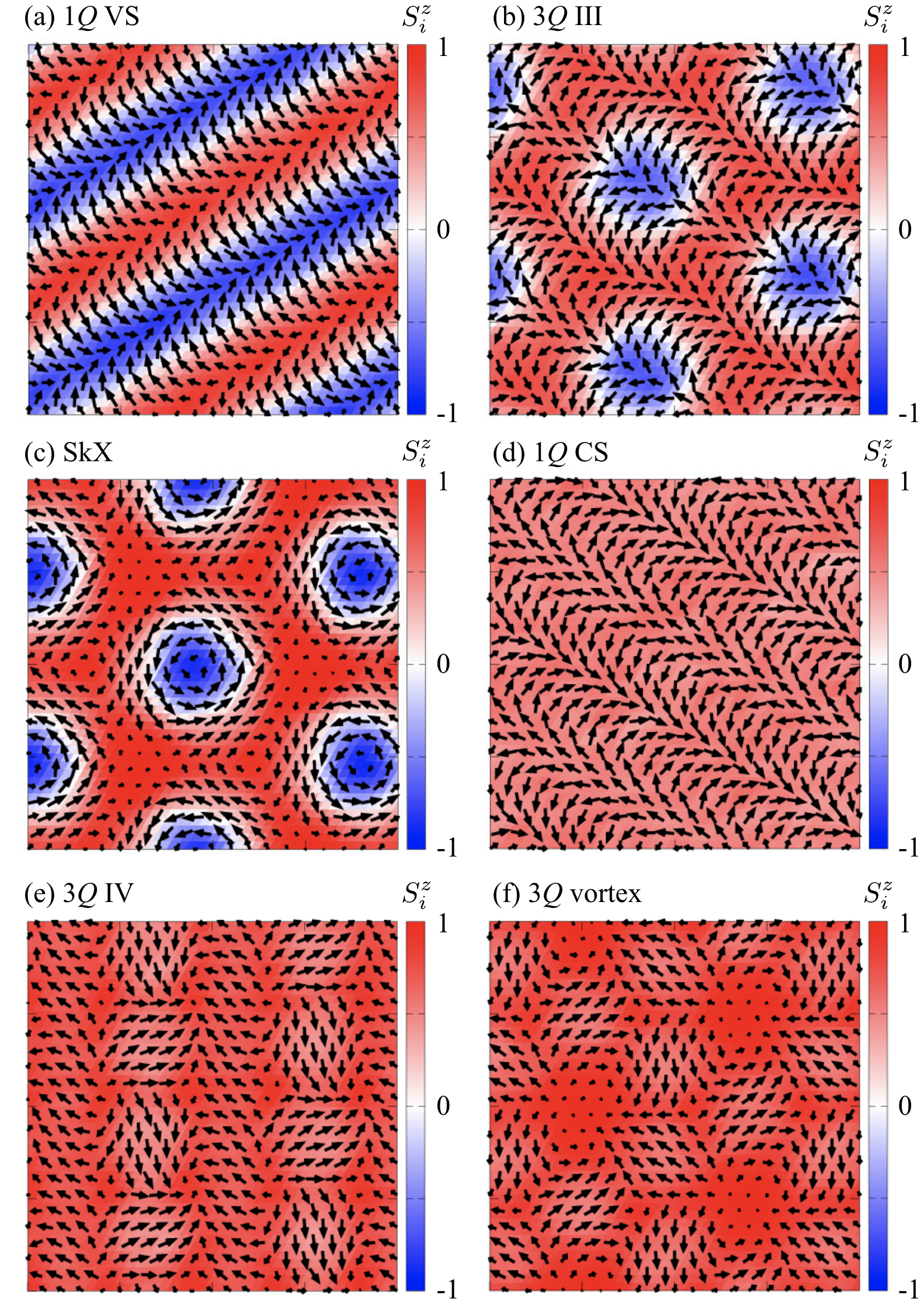} 
\caption{
\label{fig:Byzx_spin}
Real-space spin configurations of (a) the single-$Q$ vertical spiral (1$Q$ VS) state at $A_{43'}=0.3$ and $H=0.3$, (b) the triple-$Q$ (3$Q$) III state at $A_{43'}=0.3$ and $H=0.7$, (c) the SkX at $A_{43'}=0.3$ and $H=0.8$, (d) the 1$Q$ conical spiral (CS) state at $A_{43'}=0.3$ and $H=1$, (e) the 3$Q$ IV state at $A_{43'}=0.3$ and $H=1.3$, and (f) the 3$Q$ vortex state at $A_{43'}=0.3$ and $H=1.5$. 
}
\end{center}
\end{figure}

\begin{figure}[t!]
\begin{center}
\includegraphics[width=0.8 \hsize ]{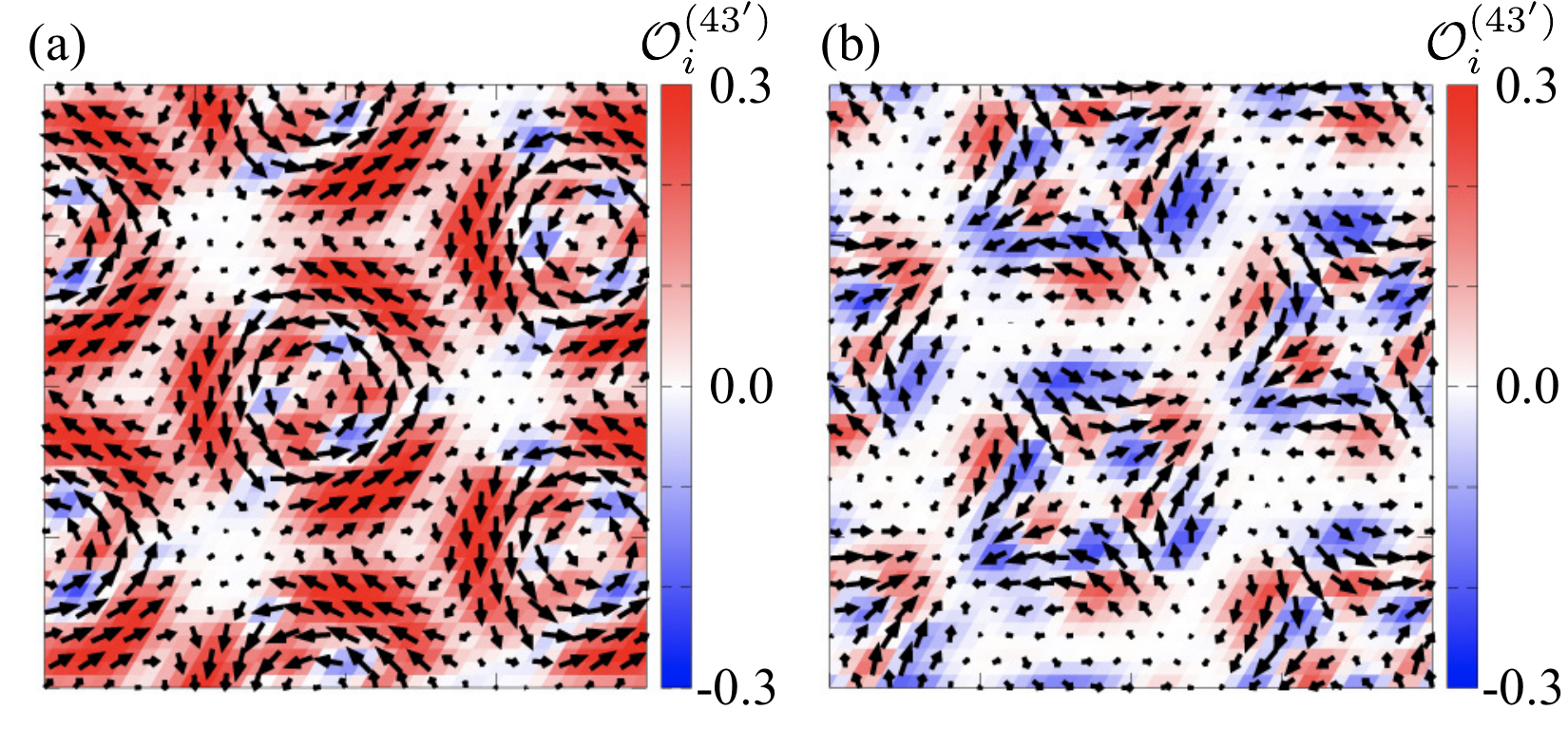} 
\caption{
\label{fig:Byzx_spin_SkX}
Contour of $\mathcal{O}^{(43')}_i$ in the SkXs corresponding to (a) Fig.~\ref{fig:Byzx_spin}(c) and (b) Fig.~\ref{fig:Bz_spin}(c). 
}
\end{center}
\end{figure}

Next, we consider the effect of the trigonal single-ion anisotropy on the SkX. 
As discussed in Sec.~\ref{sec:Model and method}, there are four types of the trigonal single-ion anisotropy, $A_{43}$, $A_{43'}$, $A_{63}$, and $A_{63'}$. 
We here present the results of $A_{43'}$ and $A_{63'}$, since the results of $A_{43}$ and $A_{63}$ are obtained by exchanging $x$ and $y$ spin components only when considering the single component of $A_{lm}$. 

Figure~\ref{fig:PD_Byzx}(a) shows the phase diagram in the plane of $A_{43'}$ and $H$ obtained by the simulated annealing. 
The same magnetic phase diagram is obtained for $A_{43'}<0$. 
Similar to the results under the uniaxial anisotropy in Sec.~\ref{sec:Uniaxial anisotropy}, the SkX is stabilized for nonzero $A_{43'}$. 
On the other hand, the instability tendency toward the other single-$Q$ and multiple-$Q$ states is different from that under the uniaxial anisotropy. 
For example, the triple-$Q$ III, triple-$Q$ IV, and triple-$Q$ vortex states appear instead of triple-$Q$ I and triple-$Q$ II states. 
Moreover, one finds that the single-$Q$ conical spiral state that appears only for the easy-plane uniaxial anisotropy in Fig.~\ref{fig:PD_Bz} is stabilized under $A_{43'}$ in the vicinity of the SkX. 

We show the phase sequence against $H$ at $A_{43'}=0.3$ in Fig.~\ref{fig:Byzx_Mq}(a). 
There are six magnetic phases except for the fully-polarized state, which are distinguished by the first-order phase transition with jumps of $M^z$, $\chi_0$, and $\bm{m}_{\bm{Q}_\nu}$
The behaviors of $M^z$ and $\bm{m}_{\bm{Q}_\nu}$ in the single-$Q$ vertical spiral state are similar to those for $A_{20}$ in Fig.~\ref{fig:Bz_Mq}(a). 
The spin configuration of the single-$Q$ vertical spiral state is shown in Fig.~\ref{fig:Byzx_spin}(a). 
The single-$Q$ vertical spiral state changes into the triple-$Q$ III state with a different triple-$Q$ superposition from the triple-$Q$ I state in Fig.~\ref{fig:Bz_Mq}(a). 
The triple-$Q$ III state exhibits the triple-$Q$ modulations in both $xy$ and $z$ spin components, as shown in Fig.~\ref{fig:Byzx_Mq}(a); $m^{xy}_{\bm{Q}_1}>m^{xy}_{\bm{Q}_2}=m^{xy}_{\bm{Q}_3}$ and $m^{z}_{\bm{Q}_1}\simeq m^{z}_{\bm{Q}_2}=m^{z}_{\bm{Q}_3}$. 
As $m^{xy}_{\bm{Q}_1}$ is dominant and the magnitudes of $m^{z}_{\bm{Q}_\eta}$ are almost equivalent, the real-space spin configuration of the triple-$Q$ III state is characterized by a superposition of the conical spiral [Fig.~\ref{fig:Byzx_spin}(d)] and the $3Q$ bubble I state [Fig.~\ref{fig:Bz_spin}(f)], as shown in Fig.~\ref{fig:Byzx_spin}(b). 

The triple-$Q$ III state is replaced by the SkX with the increase of $H$, as shown in Fig.~\ref{fig:PD_Byzx}. 
This indicates that the trigonal-type single-ion anisotropy also leads to the SkX that has not been reported so far. 
Although the real-space spin configuration of the SkX in Fig.~\ref{fig:Byzx_spin}(c) seems to be similar to that by the uniaxial anisotropy in Fig.~\ref{fig:Bz_spin}(c), the spin configuration around the skyrmion core is modulated so as to gain the energy by the single-ion anisotropy $A_{43'}$. 
To demonstrate that, we show the contour plot of $\mathcal{O}_{i}^{(43')}$ in the SkXs under $A_{43'}$ in Fig.~\ref{fig:Byzx_spin_SkX}(a) and $A_{20}$ in Fig.~\ref{fig:Byzx_spin_SkX}(b). 
Clearly, there is almost no energy gain by $A_{43'}$ in the SkX in the hexagonal system [Fig.~\ref{fig:Byzx_spin_SkX}(b)], while the SkX in the trigonal system gains the energy by $A_{43'}$ [Fig.~\ref{fig:Byzx_spin_SkX}(a)]. 
Similar to the case under the uniaxial-type single-ion anisotropy, the SkXs with the skyrmion number of $\pm 1$ (SkX and anti-type SkX), are degenerate in the presence of the trigonal-type single-ion anisotropy. 

The high-field phase of the SkX is the single-$Q$ conical state, which is different from the result in Sec.~\ref{sec:Uniaxial anisotropy}. 
The spin configuration of the single-$Q$ conical state is shown in Fig.~\ref{fig:Byzx_spin}(d).
While further increasing $H$, the triple-$Q$ IV state appears. 
This state exhibits the dominant inplane modulations along with the $\bm{Q}_1$ direction and the subdominant inplane modulations along with the $\bm{Q}_2$ and $\bm{Q}_3$ directions, as shown in the right panel of Fig.~\ref{fig:Byzx_Mq}(a). 
This state also shows a small out-of-plane spin modulation along with the $\bm{Q}_1$ direction. 
Such a feature is found in the real-space spin configuration in Fig.~\ref{fig:Byzx_spin}(e), where there is a small modulation in the $z$-spin component. 
As shown in the phase diagram in Fig.~\ref{fig:PD_Byzx}(a), the single-$Q$ conical state is also replaced by the triple-$Q$ IV state by increasing $A_{43'}$. 
At $A_{43'}=0.5$, the single-$Q$ conical state vanishes, and the direct phase transition between the SkX and the triple-$Q$ IV state appears. 

The triple-$Q$ IV state further changes into the triple-$Q$ vortex state with the triple-$Q$ inplane spiral along with the $\bm{Q}_1$, $\bm{Q}_2$, and $\bm{Q}_3$ directions with equal intensity, as shown in the right panel of Fig.~\ref{fig:Byzx_Mq}(a). 
This spin texture is regarded as a periodic alignment of the vortex with the winding number of $2$, as found in the real-space spin configuration in Fig.~\ref{fig:Byzx_spin}(f). 
Although this spin state in a high-field region is similar to the vortex crystal found in frustrated magnets~\cite{Kamiya_PhysRevX.4.011023,Hayami_PhysRevB.94.174420}, the winding number around each vortex is doubled as that in the previous studies. 
The difference is owing to the presence of the single-ion anisotropy $A_{43'}$, which favors the threefold-symmetric vortex with the winding number of two rather than the sixfold-symmetric vortex with the winding number of one~\cite{hayami2019magnetic}.
It is noted that the total winding number in a magnetic unit cell becomes zero, as there are vortices with the winding number of $-1$ and their number is double that of the winding number of $2$. 

Figure~\ref{fig:PD_Byzx}(b) shows the phase diagram under $A_{63'}$. 
The overall phase diagram is similar to that in Fig.~\ref{fig:PD_Byzx}(a). 
The behaviors of $M^z$ and $\chi_0$ are also similar to those under $A_{43'}$, as compared to the left panels of Figs.~\ref{fig:Byzx_Mq}(a) and \ref{fig:Byzx_Mq}(b). 
Thus, the trigonal single-ion anisotropy, $A_{43'}$ and $A_{63'}$, tends to favor the SkX as the uniaxial one. 
The difference between $A_{43'}$ and $A_{63'}$ is found in the appearance of the triple-$Q$ state with a weak triple-$Q$ modulation [denoted as weak 3$Q$ in Fig.~\ref{fig:PD_Byzx}(b)] in the intermediate field region. 
As shown in the right panel of Fig.~\ref{fig:Byzx_Mq}(b), the intensities of $\bm{m}_{\bm{Q}_2}$ and $\bm{m}_{\bm{Q}_3}$ are much smaller than that of $\bm{m}_{\bm{Q}_1}$. 
Reflecting the small modulation, the real-space spin texture is similar to that in the single-$Q$ conical state (not shown).

\subsection{Hexagonal inplane anisotropy}
\label{sec:Hexagonal inplane anisotropy}

\begin{figure}[t!]
\begin{center}
\includegraphics[width=0.6 \hsize ]{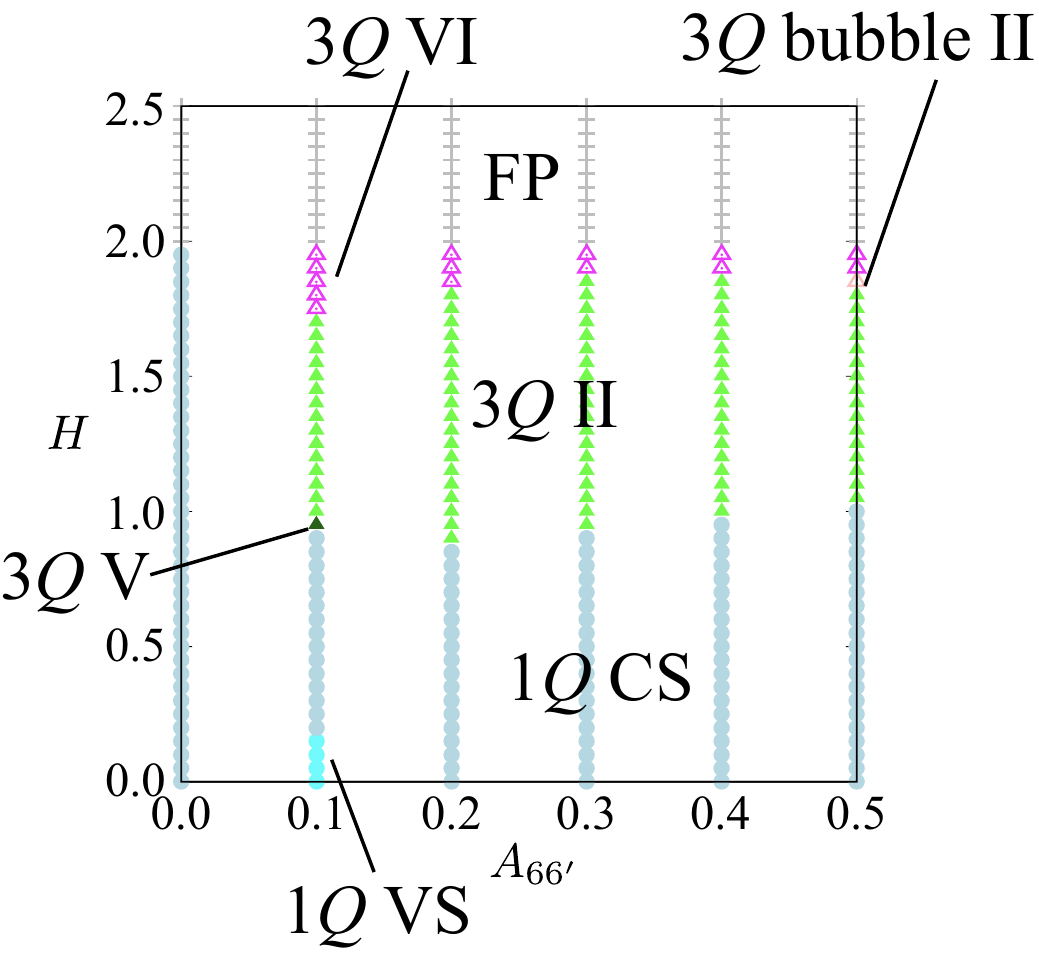} 
\caption{
\label{fig:PD_Bxy6}
Magnetic phase diagram under the hexagonal inplane single-ion anisotropy $A_{66'}$. 
The magnetic field $H$ is applied along the $z$ direction.
``VS", ``CS", and ``FP" represent the vertical spiral, conical spiral, and the fully-polarized states, respectively. 
}
\end{center}
\end{figure}

\begin{figure}[t!]
\begin{center}
\includegraphics[width=0.8 \hsize ]{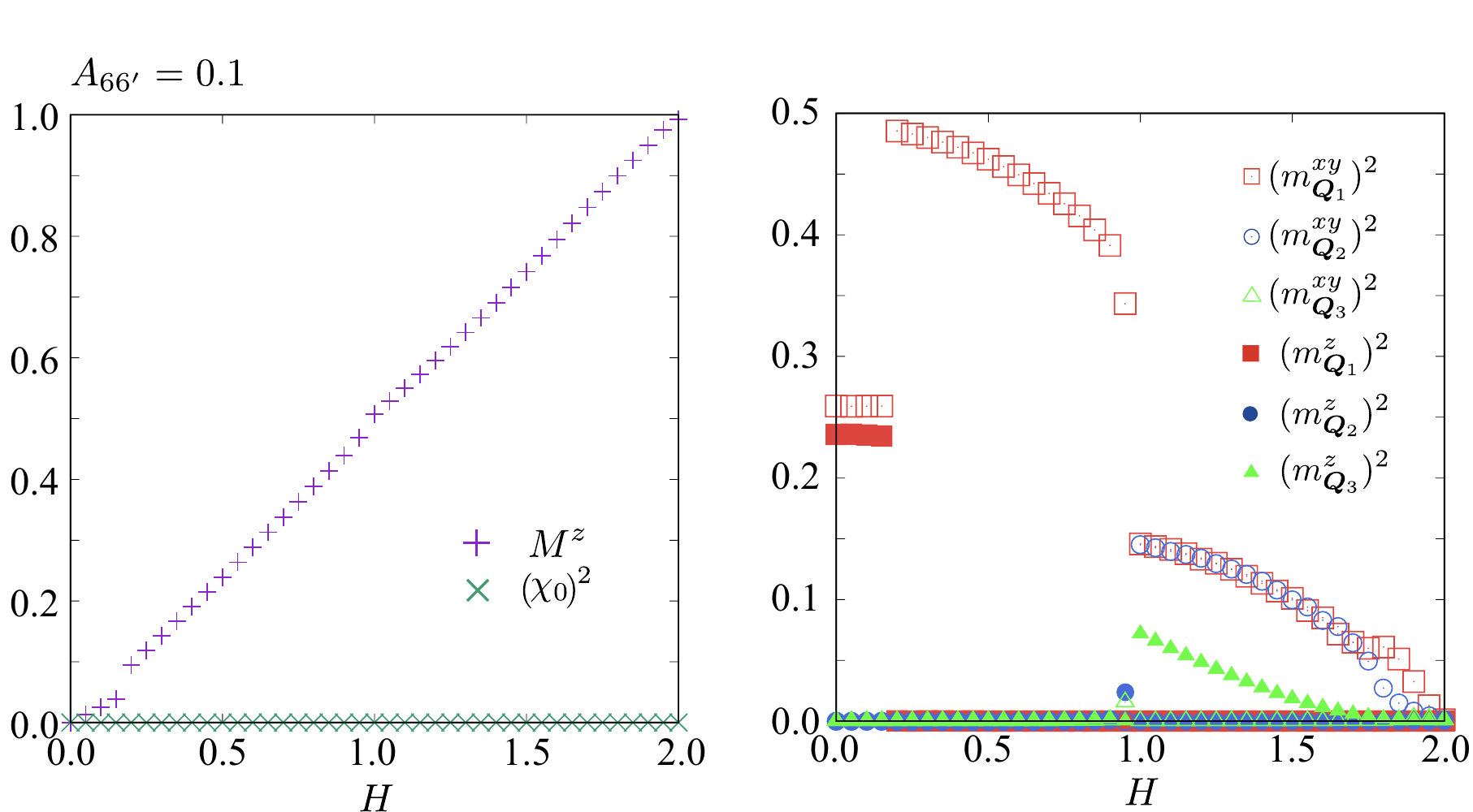} 
\caption{
\label{fig:Bxy6_Mq}
$H$ dependence of (left panel) $M^z$ and $(\chi_0)^2$ and (right panel) $(m_{\bm{Q}_\eta}^{xy})^2$ and $(m_{\bm{Q}_\eta}^{z})^2$ for $A_{66'}=0.1$. 
The magnetic field $H$ is applied along the $z$ direction.
}
\end{center}
\end{figure}

\begin{figure}[t!]
\begin{center}
\includegraphics[width=0.8 \hsize ]{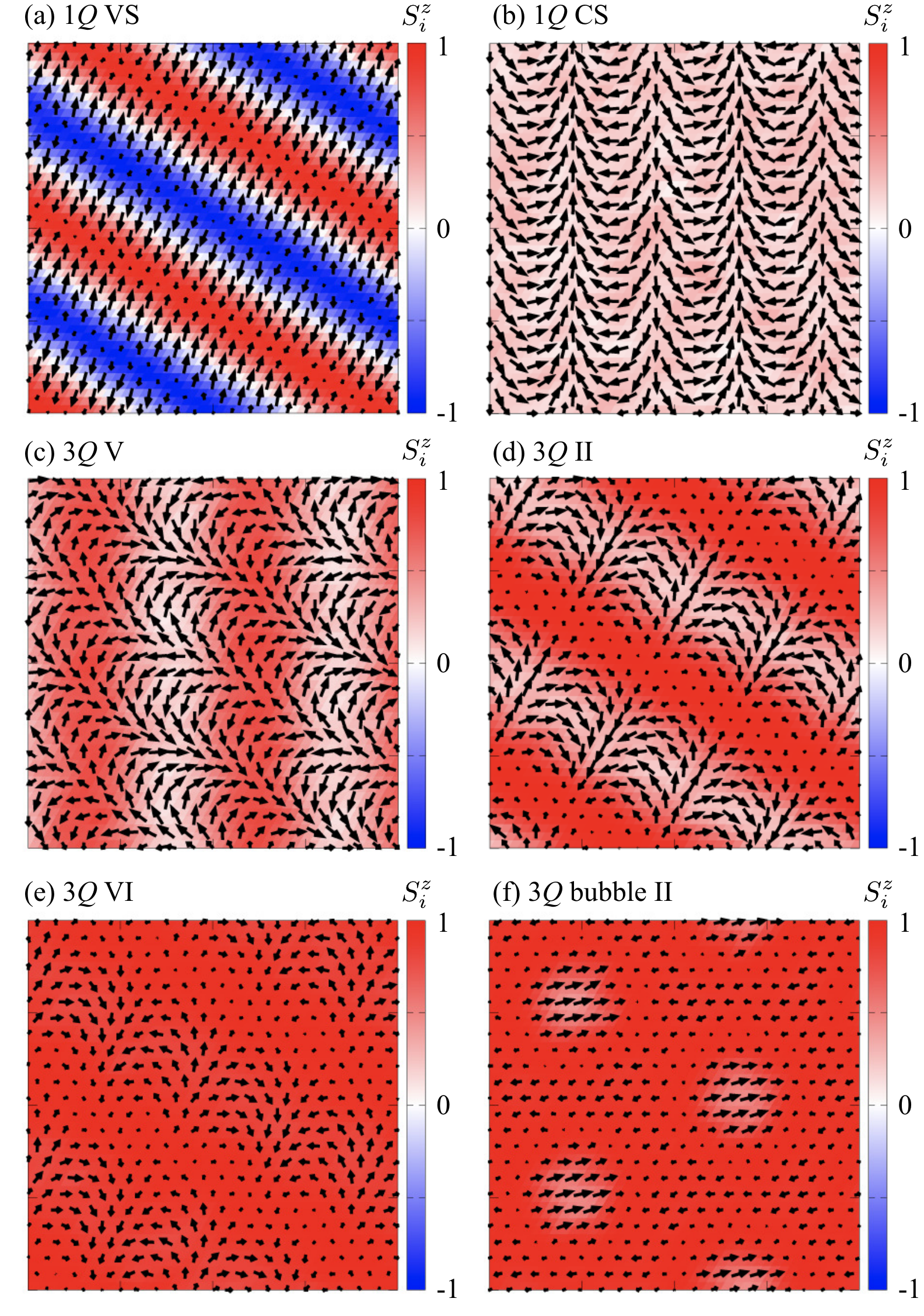} 
\caption{
\label{fig:Bxy6_spin}
Real-space spin configurations of (a) the single-$Q$ vertical spiral (1$Q$ VS) state at $A_{66'}=0.1$ and $H=0.15$, (b) the 1$Q$ conical spiral (CS) state state at $A_{66'}=0.1$ and $H=0.4$, (c) the triple-$Q$ (3$Q$) V at $A_{66'}=0.1$ and $H=0.95$, (d) the 3$Q$ II state at $A_{66'}=0.1$ and $H=1.3$, (e) the 3$Q$ VI state at $A_{66'}=0.1$ and $H=1.85$, and (f) the 3$Q$ bubble II state at $A_{66'}=0.5$ and $H=1.85$. 
}
\end{center}
\end{figure}

Finally, we consider the effect of the single-ion anisotropy, $\mathcal{O}^{(66)}_i$ and $\mathcal{O}^{(66')}_i$, which is classified into the hexagonal inplane anisotropy. 
As the difference between $\mathcal{O}^{(66)}_i$ and $\mathcal{O}^{(66')}_i$ leads to the different spin orientation and does not give a qualitatively different phase, we only examine the effect of $A_{66'}$ on the magnetic phase diagram. 
Similar to $A_{43'}$, this single-ion anisotropy also shows the symmetric phase diagram in terms of positive and negative $A_{66'}$, and hence, we focus on $A_{66'}>0$. 

Figure~\ref{fig:PD_Bxy6} shows the phase diagram when introducing $A_{66'}$. 
In contrast to the results in Secs.~\ref{sec:Uniaxial anisotropy} and \ref{sec:Trigonal anisotropy}, the SkX does not appear in the phase diagram. 
This is intuitively understood from the fact that the SkX in centrosymmetric magnets tends to be stabilized by the easy-axis anisotropy rather than the easy-plane anisotropy, as exemplified in the phase diagram with $A_{20}<0$, $A_{40}<0$, and $A_{60}<0$ in Fig.~\ref{fig:PD_Bz}. 
On the other hand, the hexagonal inplane anisotropy gives rise to a variety of multiple-$Q$ states shown in Fig.~\ref{fig:PD_Bxy6}.  
This is distinct from the case under the uniaxial easy-plane anisotropy, where only the single-$Q$ conical spiral state is stabilized. 

We show the $H$ dependence of $M^z$, $\chi_0$, and $\bm{m}_{\bm{Q}_\eta}$ at $A_{66'}=0.1$ in Fig.~\ref{fig:Bxy6_Mq}. 
There are five magnetic states in addition to the fully-polarized state. 
The single-$Q$ vertical spiral state can be stabilized in a low-field region, since the inplane anisotropy by $A_{66'}$ exhibits an angle dependence in the $xy$ plane to satisfy the sixfold rotational symmetry; the spin polarization along the $\bm{e}_1=(1,1,0)$ [$\bm{e}_2=(\cos(\pi/12),\sin(\pi/12),0)$] and their sixfold-rotational directions lose (gain) the energy. 
Indeed, the inplane spin moments in the single-$Q$ spiral state point along with the direction from $\bm{e}_2$ rotated by $\pi/3$, as shown in Fig.~\ref{fig:Bxy6_spin}(a). 
The single-$Q$ vertical spiral state vanishes for large $A_{66'}$ or large $H$, and it is replaced by the single-$Q$ conical spiral state, whose spin configuration is shown in Fig.~\ref{fig:Bxy6_spin}(b). 

While gradually increasing $H$, the optimized spin state changes from the single-$Q$ conical spiral state to the triple-$Q$ V, the triple-$Q$ II, and the triple-$Q$ VI states. 
As shown in Figs.~\ref{fig:Bxy6_spin}(c), \ref{fig:Bxy6_spin}(d), and \ref{fig:Bxy6_spin}(e), these three states are characterized by the inplane spin modulations; the triple-$Q$ V state is mainly characterized by the single-$Q$ inplane modulation, while the other triple-$Q$ II and VI states are mainly characterized by the double-$Q$ inplane modulations. 
The detail of $\bm{m}_{\bm{Q}_\eta}$ is shown in the right panel of Fig.~\ref{fig:Bxy6_Mq}. 
For large $A_{66'}$, the triple-$Q$ bubble II state appears in the narrow region between the triple-$Q$ II and triple-$Q$ IV states, as shown in Fig.~\ref{fig:PD_Bxy6}. 
The real-space spin configuration shown in Fig.~\ref{fig:Bxy6_spin}(f) is similar to the bubble crystal in Fig.~\ref{fig:Bz_spin}(f), although the spin-polarized direction is different from each other.

\begin{figure}[t!]
\begin{center}
\includegraphics[width=0.8 \hsize ]{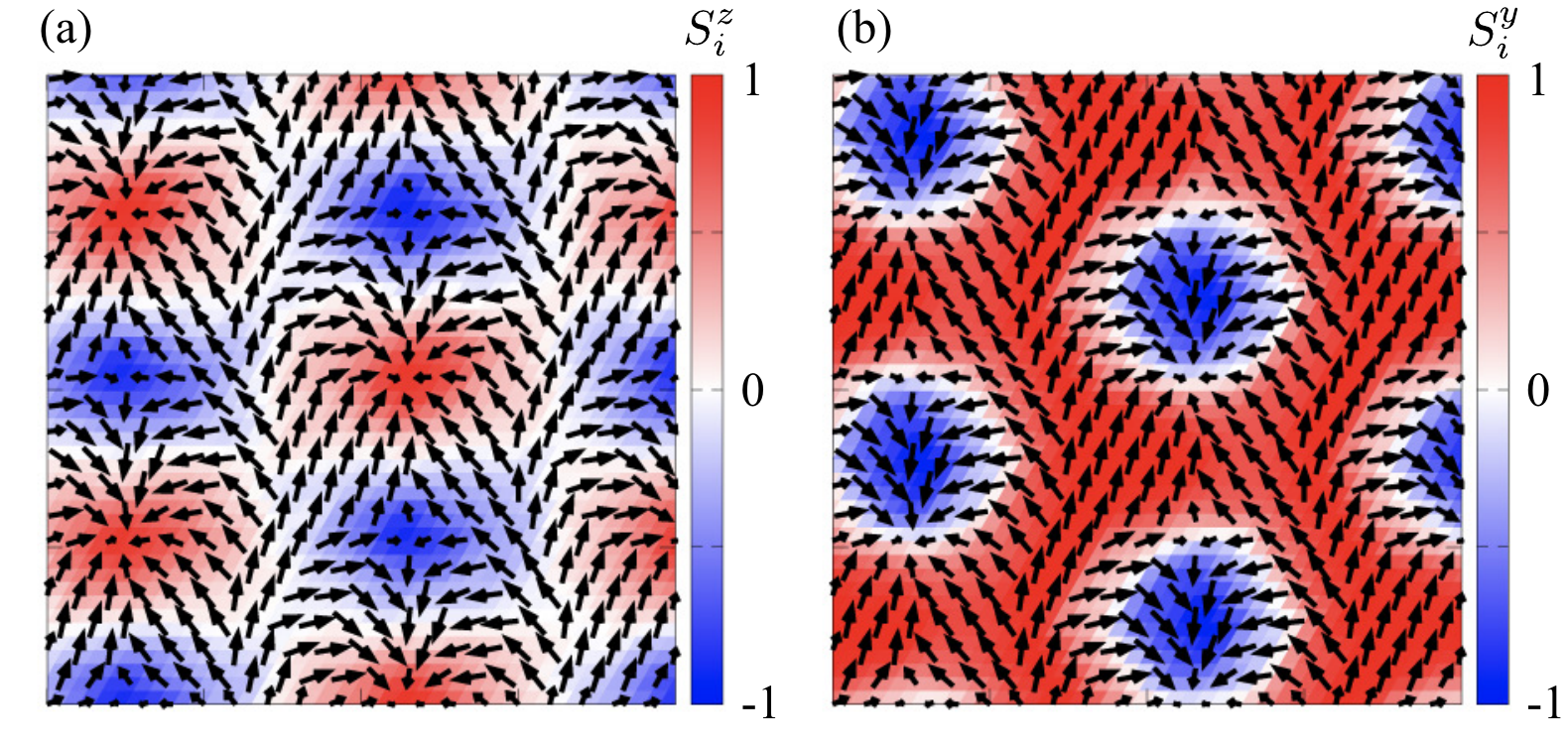} 
\caption{
\label{fig:Bxy6_spin_Hxy}
(a,b) Real-space spin configurations of the SkX at $A_{66'}=0.3$ and $H^{y}=0.8$. 
In (a) [(b)], the contour shows $S_i^z$ ($S_i^y$). 
}
\end{center}
\end{figure}

Although the SkX under the hexagonal inplane single-ion anisotropy is not stabilized in the magnetic field along the $z$ direction, it can appear in the inplane field~\cite{Hayami_PhysRevB.103.224418,Hayami_PhysRevB.103.054422}. 
We here apply the magnetic field along the $y$ direction instead of that along the $z$ direction in the form of $-H^y\sum_i S_i^y$ with $H^y=0.8$. 
Then, one finds the SkX for nonzero $A_{66'}$, where the spin configuration obtained by the simulated annealing is shown in Fig.~\ref{fig:Bxy6_spin_Hxy}. 
The spin configuration in Fig.~\ref{fig:Bxy6_spin_Hxy}(a) indicates a periodic alignment of two types of antimeron with a half skyrmion number $+1/2$: One is characterized by the vortex with $S_i^z>0$ and the winding number $+1$ and the other is the vortex with $S_i^z<0$ and the winding number $-1$. 
In other words, the total skyrmion number in the unit cell is $+1$. 
The SkX spin texture is also understood from the contour of $S_i^y$, which corresponds to the field-direction component, in Fig.~\ref{fig:Bxy6_spin_Hxy}(b); the skyrmion core forms the triangular lattice as found in Fig.~\ref{fig:Bz_spin}(c). 
Thus, there is a chance of inducing the SkX by applying the inplane field to the hexagonal magnets with hexagonal-type easy-plane anisotropy.

\section{Summary}
\label{sec:Summary}
To summarize, we have investigated the instability toward the SkX in the presence of various types of single-ion anisotropy under the centrosymmetric hexagonal and trigonal point groups. 
By performing the simulated annealing for the triangular-lattice spin model, we found that the SkX is stabilized in the out-of-plane field under the uniaxial and trigonal anisotropy, while it is stabilized in the inplane field under the hexagonal inplane anisotropy. 
We also showed that the single-ion anisotropy gives rise to a plethora of multiple-$Q$ states depending on the types of anisotropy. 
Our results provide another way of stabilizing the SkX in centrosymmetric magnets based on single-ion anisotropy. 
In particular, there are three important key factors to stabilize the SkX: the competing exchange interaction leading to the spiral spin modulation with a finite ordering vector at zero fields, the hexagonal and trigonal crystal symmetry, and the presence of the single-ion anisotropy.
For the last condition, the materials consisting of magnetic ions with large orbital angular momentum, such as an $f$-orbital wave function, are appropriate candidates to exhibit the SkX.

In addition, the present results give a possibility of realizing further exotic multiple-$Q$ states other than the SkXs, such as the hedgehog lattices~\cite{tanigaki2015real,kanazawa2017noncentrosymmetric,fujishiro2019topological,Kanazawa_PhysRevLett.125.137202,grytsiuk2020topological,Ishiwata_PhysRevB.101.134406,Okumura_PhysRevB.101.144416,okumura2020tracing,Mendive-Tapia_PhysRevB.103.024410,Shimizu_PhysRevB.103.054427,Kato_PhysRevB.104.224405}, meron-antimeron crystal~\cite{Lin_PhysRevB.91.224407,yu2018transformation,hayami2018multiple,kurumaji2019skyrmion,Hayami_PhysRevB.104.094425,chen2022triple}, vortex crystal~\cite{Momoi_PhysRevLett.79.2081,Kamiya_PhysRevX.4.011023,Wang_PhysRevLett.115.107201,Marmorini2014,Hayami_PhysRevB.94.174420,Solenov_PhysRevLett.108.096403,Ozawa_doi:10.7566/JPSJ.85.103703,takagi2018multiple,yambe2020double}, and bubble state~\cite{lin1973bubble,Garel_PhysRevB.26.325,takao1983study,Hayami_PhysRevB.93.184413,Su_PhysRevResearch.2.013160,seo2021spin}. 
Indeed, we found instabilities toward the vortex and bubble crystals in the presence of different types of single-ion anisotropy, as shown in Figs.~\ref{fig:Byzx_spin}(f) and \ref{fig:Bxy6_spin}(f). 
As different spin and scalar chirality configurations emerge depending on the resultant multiple-$Q$ superposition, it is expected to obtain intriguing electronic states~\cite{Christensen_PhysRevX.8.041022,Hayami_PhysRevB.104.144404,Hayami_PhysRevB.105.024413}, excitation spectra~\cite{Mochizuki_PhysRevLett.114.197203,mochizuki2015dynamical,garst2017collective,Santos_PhysRevB.97.024431,Weber_PhysRevB.97.224403,Kato_PhysRevB.104.224405}, and nonreciprocal nonlinear transport~\cite{Hayami_PhysRevB.101.220403,Hayami_PhysRevB.102.144441,hayami2021phase} driven by the magnetic phase transition. 
The present study provides a platform to study such phenomena based on the microscopic model.

This research was supported by JSPS KAKENHI Grants Numbers JP19K03752, JP19H01834, JP21H01037, and by JST PRESTO (JPMJPR20L8).
Parts of the numerical calculations were performed in the supercomputing systems in ISSP, the University of Tokyo.

\bibliography{ref}

\end{document}